# PPMXL photometric study of four open cluster candidates
## *(Ivanov 2, Ivanov 7, Ivanov 9 and Harvard 9)*

**Tadross, A. L. & Bendary, R.**

*National Research Institute of Astronomy and Geophysics, Egypt*

Email: altadross@yahoo.com

## ABSTRACT

The astrophysical parameters of four unstudied open star cluster candidates - Harvard 9, Ivanov 2, Ivanov 7, and Ivanov 9 - have been estimated for the first time using the PPMXL database. The stellar density distributions and color-magnitude diagrams for each cluster are used to determine the geometrical structure (cluster center, limited radius, core and tidal radii, the distances from the Sun, from the Galactic center and from the Galactic plane). Also, the main photometric parameters (age, distance modulus, color excesses, membership, total mass, relaxation time, luminosity and mass functions) are estimated.

## 1. Introduction

Open star clusters (OCs) are important celestial bodies in understanding star formation, initial mass function and stellar evolution theories. Gaburov and Gieles (2008) provided statistically significant samples of star clusters of known distance, age and metallicity. Color-magnitude Diagram (CMD) analysis through isochrones give us good estimations about the astrophysical parameters of the clusters, e.g. age, reddening and distance. In the last decades, many studies have been performed using different techniques; started from photographic photometry to the couple charge devices, CCD-photometry, and finally employing many isochrones models. The large amount of results has produced in the literature and gathered in catalogs and databases, e.g. Webda and Dias. The present study depends mainly on the PPMXL database of Röser et al. (2010). The most important thing for using PPMXL database lies in containing the positions, proper motions of USNO-B1.02 and the Near Infrared (NIR) photometry of the Two Micron All Sky Survey (2MASS), which allowed us to search and study clusters within the disk of the Galaxy that is normally obscured by dust and gas clouds. Such databases help us to unravel star formation problems and answer questions about the spiral structure of the Milky Way Galaxy.

Our candidates are selected from among the unstudied OCs listed in Dias catalog. Fig. 1 represents the clusters' images as taken from the LEDAS DSS image of Digitized Sky Surveys (http://www.ledas.ac.uk/). The only available information known about these clusters are the coordinates and the apparent diameters in arc minutes. Equatorial and Galactic positions of our candidates with their apparent diameters are listed in Table 1. The quality of the downloaded data is taken into account and the estimated physical properties of each cluster are achieved by applying the same methodology.

*Table 1: Equatorial, Galactic coordinates and the apparent diameters of the candidate clusters.*

| Cluster | Ra.$^{h\ m\ s}$ | Dec.$^{o\ '\ ''}$ | G. Long.$^{o}$ | G. Lat.$^{o}$ | D$'$ |
|---|---|---|---|---|---|
| Ivanov 2 | 06 15 53 | +14 16 00 | 196.214 | -1.198 | 2 |
| Ivanov 9 | 06 59 43 | -04 04 00 | 217.494 | -0.016 | 1 |
| Ivanov 7 | 07 30 40 | -15 18 00 | 230.983 | 1.484 | 2.8 |
| Harvard 9 | 15 33 46 | -53 40 00 | 325.616 | 1.899 | 6 |

This paper is organized as follows. PPMXL data extraction is presented in Section 2, while the data analysis and parameter estimations are described in Sections 3. Finally, the conclusion of our study is devoted to Section 4.

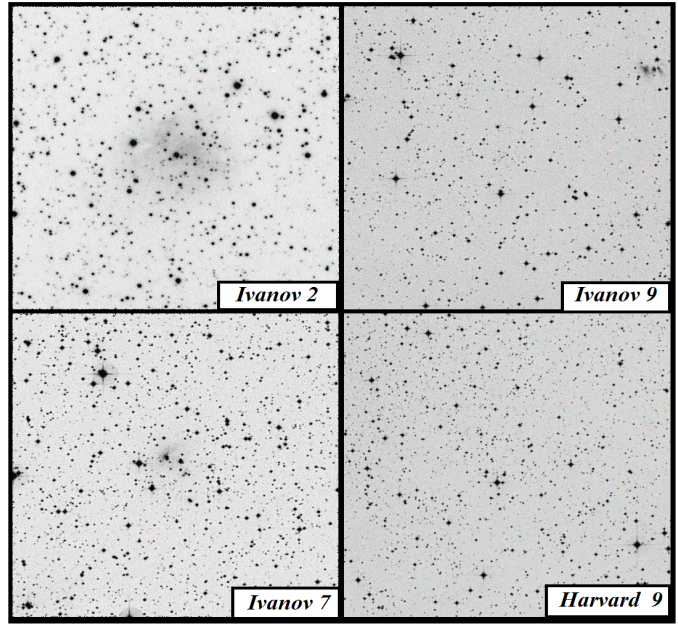

*Figure 1: The images of candidate clusters, as taken from the LEDAS DSS Digitized Sky Surveys.*

## 2. PPMXL Data Extraction

The main data of this work has been extracted from the PPMXL catalog of Röser et al. (2010). This catalog combines the USNO-B1.0 database of Monet et al. (2003) and the 2MASS database of Skrutskie et al. (2006) yielding the largest collection of proper motions in the International Celestial Reference Frame (ICRF) to date (Röser et al. 2010). USNO-B1.0 contains the positions of more than one billion objects taken photographically around 1960; 2MASS is an all-sky survey conducted in the years 1997 to 2001 in the J, H and Ks bands. In PPMXL, data from USNO-B1.0 are used as the first epoch images and those from 2MASS as the second epoch images, deriving the mean positions and proper motions for all objects of brightest magnitudes down to V ~ 20 mag. Mean errors of the proper motions vary from ~ 4 mas/yr (milliseconds per year) for J < 10 mag to more than ~10 mas/yr at J > 16 mag (Khalaj & Baumgardt 2013). Röser et al. (2010) presented the PPMXL catalog which contains ~ 910 million objects, of which 410 million objects have photometric data from 2MASS catalog.

USNO-B1.0 is a very useful catalog, which gives us an opportunity to distinguish between the members and background/field stars. On the other hand, NIR 2MASS photometry provides J (1.25μm), H (1.65μm), and Ks (2.17μm) band photometry of millions of galaxies and nearly a half-billion stars (Carpenter, 2001).

Before using the proper motion data in the PPMXL catalog, it is very important to understand the systematic and random errors of the proper motions in this catalog. Röser et al. (2010) have corrected the plate-dependent distortions of the proper motions during the construction procedure of the PPMXL catalog. They have pointed out that the magnitude and color dependent systematic errors in the PPMXL catalog are difficult to be determined due to no independent reference on ICRS at fainter magnitude (Wu, Zhen-Yu et al. 2011).

This survey has proven to be a powerful tool in the analysis of the structure and stellar content of open clusters (Bica et al. 2003, Bonatto & Bica 2003). The photometric uncertainty of the 2MASS data is less than 0.155 at Ks ≈ 16.5 mag, which is the photometric completeness for stars with $|b| > 25°$, Skrutskie et al. (2006).

It is noted that the candidate clusters are located near the Galactic plane (|b| < 2º), therefore we expect significant foreground and background field star contamination. These clusters have medium central concentration, as appeared from their images on the Digitized Sky Survey (DSS), see Fig. 1. Their apparent diameters are less than 10 arcmin, hence the downloaded data have been extended to reach the field background stars, whereas the clusters dissolved there, i.e. the data are extracted at such sizes of about 10-15 arcmin.

In this context, to get a net worksheet data for investigating clusters, the photometric completeness limit has been applied to the photometric pass-band 2MASS data to avoid the over-sampling of the lower parts of the cluster's CMDs (cf. Bonatto et al. 2004). The stars with observational uncertainties ≥ 0.20 mag have been removed. Pm vector point diagram (VPD) with distribution histogram of 2 mas/yr bins for (pm α cos δ) and (pm δ) have been constructed as shown in Fig. 2. The Gaussian function fit to the central bins provides the mean pm in both directions. All data lie at that mean ±1 σ (where σ is the standard deviation of the mean) can be considered as probable members. In addition, the stellar photometric membership criteria are adopted based on the location of the stars within ±0.1 mag around the zero age main sequence (ZAMS) curves in the CMDs, (Clariá & Lapasset 1986).

## 3. Data Analysis
## 3.1. Cluster's Centers and Radial Density Profile

The star-count of the candidate clusters has been applied to the cluster's area up to 10 arcmin for each of the adopted centers. Each area is divided into equal sized bins in right ascension and declination (α and δ). The purpose of this counting process is to determine the maximum central density of the clusters. The clusters' centers are found by fitting Gaussian distribution function to the profiles of star counts in α and δ respectively, as shown in Fig. 3. The cluster center is defined as the location of the maximum stellar density of the cluster's area. The differences between our estimated center and the obtained one of Webda are shown in Table 2.

*Table 2: The differences between our estimated center and the obtained one of Webda Catalog.*

| Cluster | Webda | | Present study | |
|---|---|---|---|---|
| | Ra. $^{h\ m\ s}$ | Dec. $^{o\ \prime\ \prime}$ | Ra. $^{h\ m\ s}$ | Dec. $^{o\ \prime\ \prime}$ |
| Ivanov 2 | 06 15 53 | +14 16 00 | 06 15 48 | +14 15 51 |
| Ivanov 9 | 06 59 43 | -04 04 00 | 06 59 47 | -04 04 00 |
| Ivanov 7 | 07 30 40 | -15 18 00 | 07 30 39 | -15 18 04 |
| Harvard 9 | 15 33 46 | -53 40 00 | 15 33 44 | -53 34 57 |

To establish the radial density profile (RDP) of the clusters under consideration, their areas are divided into central concentric circles with bin sizes $R_i \leq 1$ arcmin, started from the cluster center. The number density, $R_i$, in the i$^{th}$ zone is calculated by using the formula of $R_i = N_i / A_i$. Where $N_i$ is the number of stars and $A_i$ is the area of the i$^{th}$ zone. The star counts of the next steps should be subtracted from the previous ones, so that we obtained only the amount of the stars within the relevant shell's area, not a cumulative count. The density uncertainties in each shell were calculated using Poisson noise statistics. Finally, we applied the empirical King model (1966), where it model parameterizes the density function $\rho(r)$ as:

$$\rho(r) = f_{bg} + \frac{f_0}{1 + (r/r_c)^2}$$

Where $f_{bg}$, $f_o$ and $r_c$ are background, central star density and the core radius of the cluster respectively. The cluster's limiting radius can be defined at that radius which covers the entire cluster area and reaches enough stability with the background field density. It is noted that PPMXL catalog allows us to obtain reliable data on the projected distribution of stars for large extensions to the clusters' halos. We can infer that open clusters appear to be somewhat larger in the near-infrared than in the optical data, Sharma et al. (2006). Because of strong field stars contamination, it is not possible to completely separate all field stars from cluster members as shown for Ivanov 9 and Harvard 9. The limiting radius of a cluster can be described with an observational border, which depends on the spatial distribution of stars in the cluster and the density of the membership and the degree of field-star contamination. Fig. 4 shows the RDP from the new centers of the clusters under consideration. The core radius and the background field density are estimated and shown for each cluster as well.

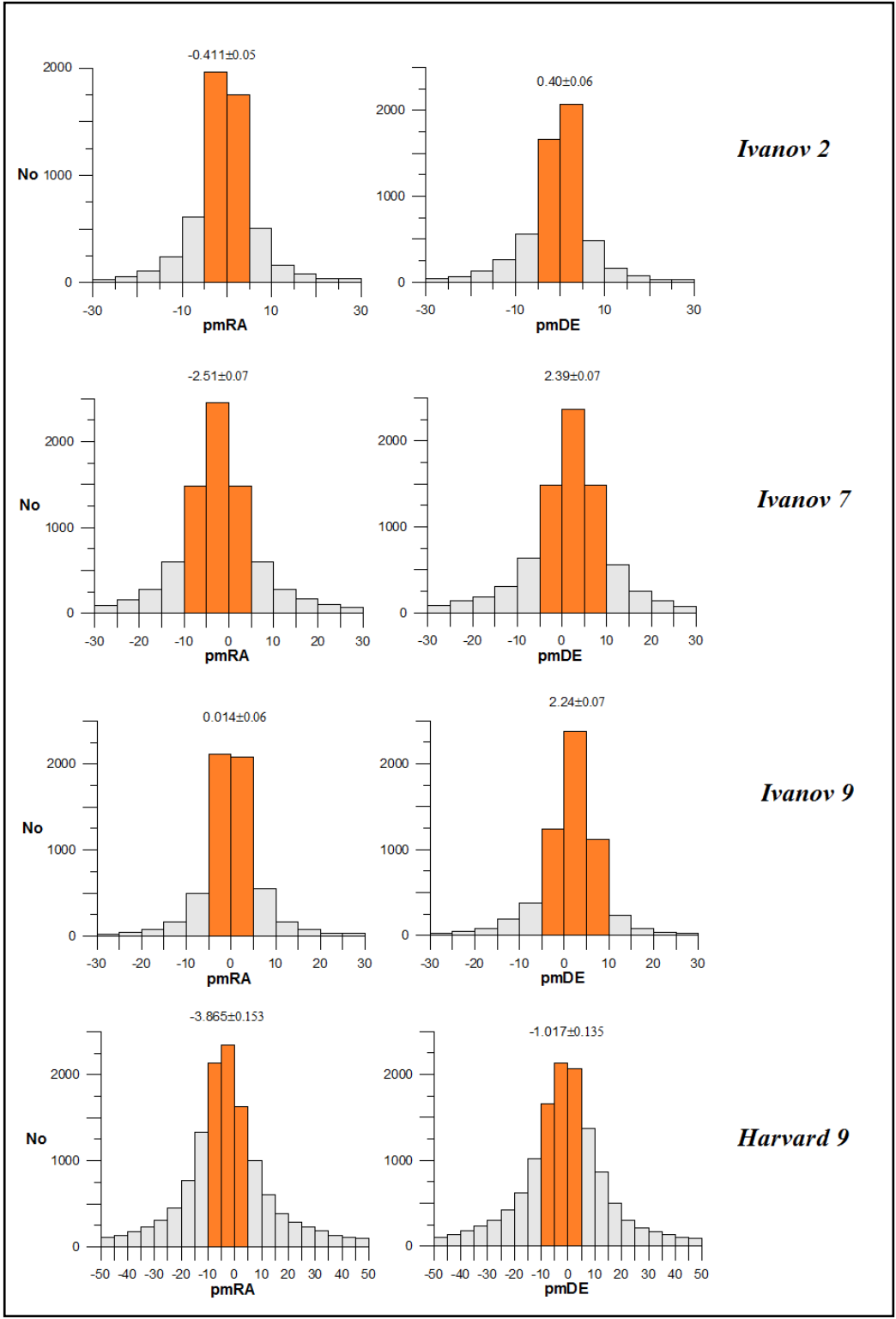

*Figure 2: Proper motion histograms of 2 mas/yr bins in right ascension and declination of the candidate clusters, the pm errors ≥ 4 mas/yr are excluded. The Gaussian function fit to the central bins provides the mean values in both directions as shown in each panel.*

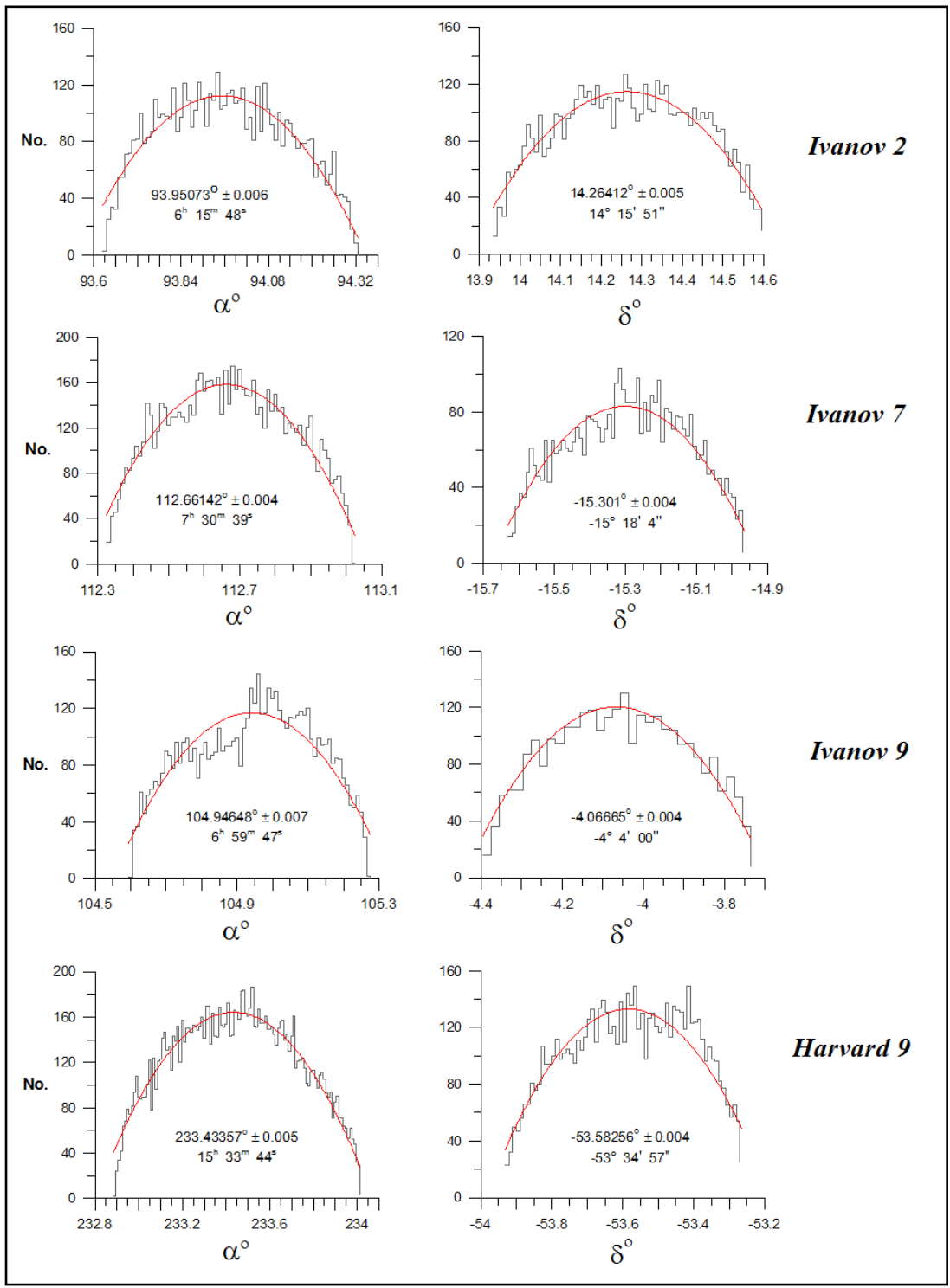

*Figure 3: The cluster center estimating of the candidate clusters. The Gaussian fit provides the coordinates of highest density areas in α and δ for each cluster respectively. The center of symmetry about the peaks of α and δ is taken to be the position of the clusters' centers.*

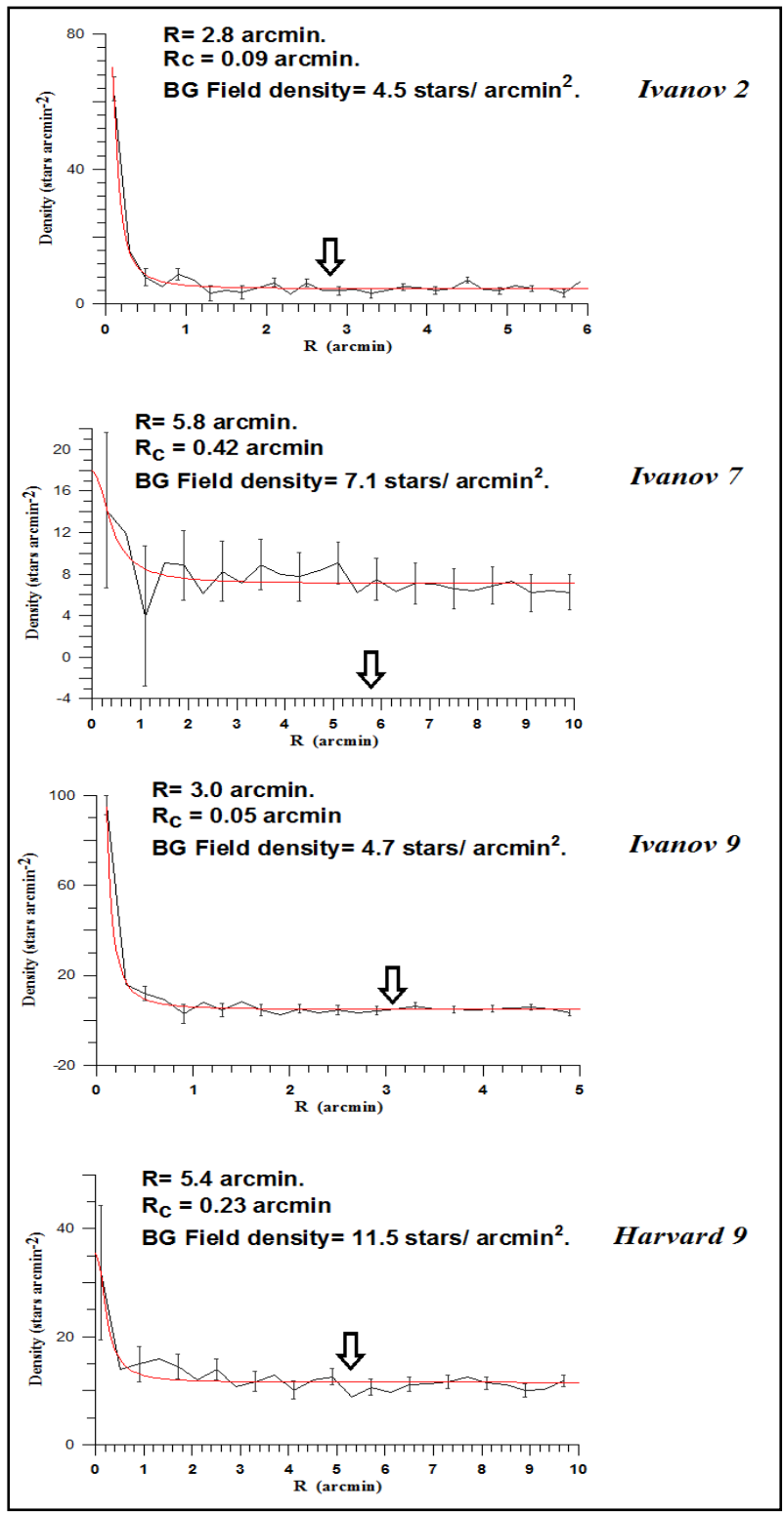

*Figure 4: The radial density profile of the clusters under consideration. Clusters' limiting radius, core radius and the background field density are estimated and shown for each candidate clusters. The curved solid line represents the fitting of King (1966) model. Error bars are determined from sampling statistics $[1/(N)^{0.5}$ where N is the number of stars used in the density estimation at that point].*

Finally, knowing the cluster's total mass (Sec. 3.4), the tidal radius can be calculated by applying the equation of Jeffries et al. (2001):

$$R_t = 1.46 \, (M_c)^{1/3}$$

Where $R_t$ and $M_c$ are the tidal radius and the total mass of the candidate clusters respectively. Nilakshi et al. (2002) noted that the halos' sizes are smaller for older systems. On the other hand, we can infer that open clusters appear to be somewhat larger in the near-infrared than in the optical data, Sharma et al. (2006).

## 3.2. The PPMXL Photometry

Depending on the PPMXL data, deep stellar analyses of the candidate clusters have been presented. The photometric data of PPMXL not only allow us to construct of relatively well defined CM diagrams of the clusters, but also permit a more reliable determination of astrophysical parameters. In this paper, we used extraction areas having a radius of 10 arcmin, which are larger than the estimated limiting radius of the clusters. Because of the weak contrast between the cluster and the background field density, some inaccurate statistical results may be produced beyond the real limit of cluster borders (Tadross, 2005).

The main astrophysical parameters of the clusters, e.g. age, reddening, distance modulus, can be determined by fitting the isochrones to the cluster CMDs. To do this, we applied several fittings on the CMDs of the clusters by using the stellar evolution models of Marigo et al. (2008) and Girardi et al. (2010) of Padova isochrones on the solar metallicity. It is worth mentioning that the assumptions of solar metallicity are quite adequate for young and intermediate age open clusters, which are closed to the Galactic disk. So, Near-Infrared surveys are very useful for the investigation of such clusters. It is relatively less affected by high reddening from the Galactic plane. However, for a specific age isochrones, the fit should be obtained at the same distance modulus for both diagrams [J-(J-H) & $K_s$-(J-$K_s$)], and the color excesses should be obeyed Fiorucci & Munari (2003)'s relations for normal interstellar medium as shown in Fig. 5. We note that, it is difficult to obtain accurate determinations of the astrophysical parameters due to the weak contrast between clusters and field stars. So, some ranges in age estimations are given.

It might be mainly explained by the theoretical differences in adopting isochrones. Therefore, the clusters' ages are found to be (5-10); (400-800); (80-120); (200-400) Myr for Ivanov 2, Ivanov 7, Ivanov 9, and Harvard 9 respectively.

Reddening determination is one of the major steps in the cluster compilation. Therefore, it estimated guiding by Schlegel et al. (1998) and Schlafly et al. (2011) in our estimations. In this context, for color excesses transformations, we used the relation $A_J/A_V = 0.276$, $A_H/A_V = 0.176$, $A_{Ks}/A_V = 0.118$ (Dutra et al. 2002). According to Fiorucci & Munari (2003), the color excess values can be estimated with the following results:
$E(J-H)/E(B-V)=0.309\pm0.130$, $E(J-Ks)/E(B-V)=0.485\pm0.150$, assuming a constant total to selective absorption ratio $R_V=A_V/E(B-V)= 3.1$. Finally, we de-reddened the distance modulus using these formulae: $A_J/E(B-V)= 0.887$, $A_{Ks}/E(B-V)= 0.322$, then the distance of each cluster from the Sun $R_\odot$ can be calculated.

Under the assumption of $R_{gc\odot}= 7.2 \pm 0.03$ kpc of Bica & Bonatto (2006) which is based on updating the parameters of globular clusters, the estimated distances from the Galactic center, $R_{gc}$, are estimated for each cluster. Also, the distance from the Galactic plane ($Z_\odot$), and the projected distances in the Galactic plane from the Sun ($X_\odot$ & $Y_\odot$) can be determined, see Table 3. For more details about the distance calculations, see Tadross (2011).

## 3.3. Luminosity functions

It is difficult to determine the membership of a cluster using only the stellar RDP. It might be claimed that most of the stars in the inner concentric rings are quite likely members, whereas the external rings are more intensely contaminated by field stars. Therefore, the stars, which are closed to the cluster's center and near to the main-sequence (MS) in CMDs are taken to be the stellar membership of the clusters. These MS stars are very important in determining the luminosity, mass functions and the total mass of the investigated clusters. The measurements of the number of stars in a cluster with a given color and magnitude ranges are very important to understand the characteristic properties of the evolutionary stages of these objects.

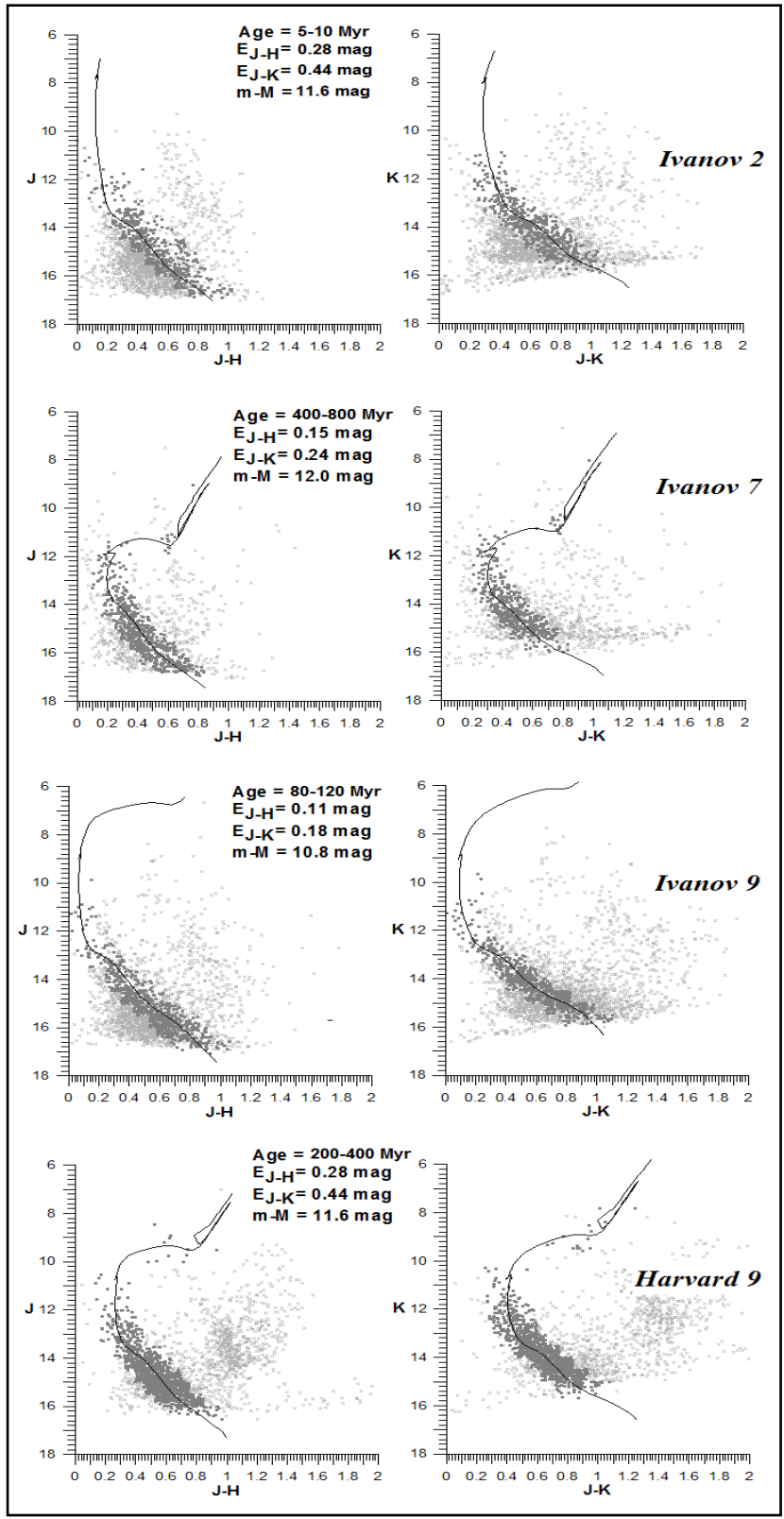

*Figure 5: The CMDs of the clusters under investigation. Dark dots represent the probable member stars lying closely to the fitted isochrones curves and light dots refer to the contaminated field stars. Age, color excess and distance modulus are shown in each cluster.*

For this purpose, we obtained the Luminosity Functions (LFs) of the four clusters by summing up the J band luminosities of all stars within the determined limiting radii. Before building the LFs, we converted the apparent J band magnitudes of possible member stars into the absolute magnitude values using the distance moduli of the clusters. We constructed the histogram sizes of LFs to include a reasonable number of stars in each absolute J magnitude bins for the best counting statistics. The total LFs of the cluster are found to be -3.65, -6.30, -3.90 and -5.20 mag for Ivanov 2, Ivanov 7, Ivanov 9, and Harvard 9 respectively; see Fig. 6.

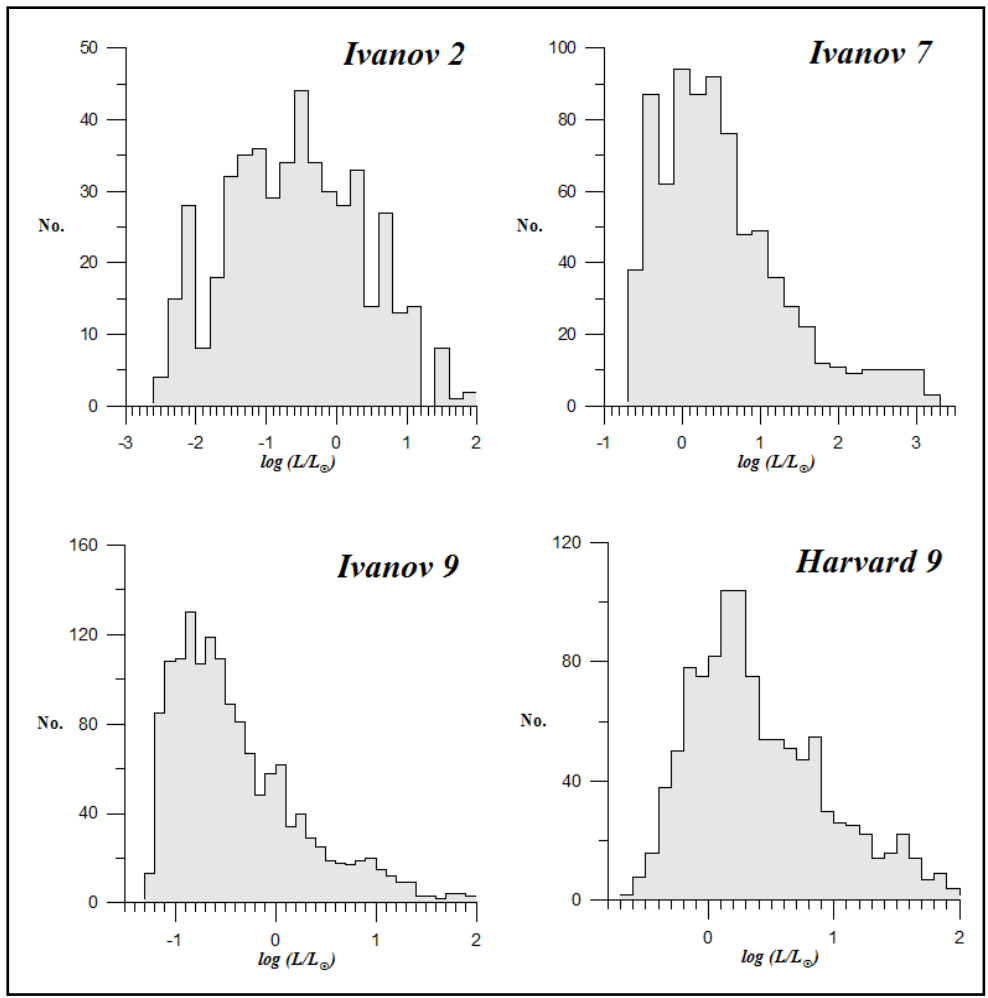

*Figure 6: The luminosity functions of the clusters under consideration.*

## 3.4. Mass functions, total masses and dynamical states

The mass functions (MFs) of the four clusters are built using the theoretical evolutionary tracks and their isochrones with different ages. The masses of possible cluster members were derived from the polynomial expression developed by Marigo et al. (2008) and Girardi et al. (2010) with solar metallicity.

The LF and MF are correlated to each other according the known Mass-luminosity relation. The accurate determination of both of them (LF & MF) suffers from the field star contamination, membership uncertainty, and mass segregation, which may affect even poorly populated, relatively young clusters (Scalo 1998). On the other hand, the properties and evolution of a star are closely related to its mass, so the determination of the initial mass function (IMF) is needed. It is an important diagnostic tool for studying large quantities of star clusters. IMF is an empirical relation that describes the mass distribution of a population of stars in terms of their theoretical initial mass. The IMF is defined in terms of a power law as follows:

$$\frac{dN}{dM} \propto M^{-\alpha}$$

Where dN/dM is the number of stars of mass interval (M:M+dM), and $\alpha$ is a dimensionless exponent. The IMF, for massive stars *(> 1 $M_\odot$)* has been studied and well established by Salpeter (1955), where $\alpha = 2.35$. This form of Salpeter shows that the number of stars in each mass range decreases rapidly with increasing mass. It is noted that the investigated MF slope ranging of the clusters under consideration are found to be -2.47, -2.73, -5.10 and -2.31, which are found to be around the Salpeter's value as shown in Fig. 7.

To estimate the total mass of the candidate clusters, the mass of each star has been estimated from a polynomial equation developed from the data of the solar metallicity isochrones (absolute magnitudes versus actual masses) at the ages of the clusters. The summation of multiplying the number of stars in each bin by the mean mass of that bin yields the total mass of the clusters, which are found to be 580, 470, 645 and 1450 $M_\odot$ for Ivanov 2, Ivanov 7, Ivanov 9, and Harvard 9 respectively.

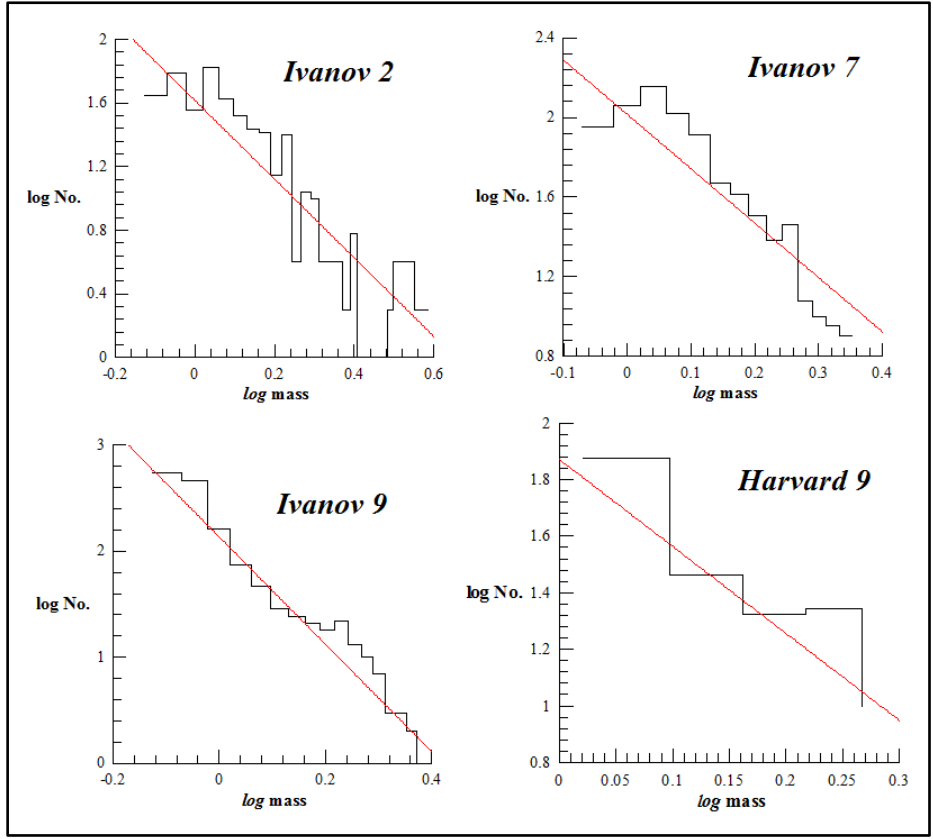

*Figure 7: The mass functions of the clusters under consideration.*

## 4. Conclusions

In the present work, we have analysed the unstudied four open clusters Ivanov 2, Ivanov 7, Ivanov 9, and Harvard 9. The astrophysical parameters have been estimated for the first time for these clusters using the PPMXL database of Röser et al. (2010). The main conclusions are summarized and listed in Table 3.

## Acknowledgments

We thank the anonymous referee for her/his valuable comments and suggestions. It is worthy to mention that, this publication made use of WEBDA, DIAS catalogs, and the data products from the PPMXL database of Röser et al. (2010).

*Table 3: The astrophysical parameters of the candidate clusters.*

| Parameter | Cluster name | | | |
|---|---|---|---|---|
| | Ivanov 2 | Ivanov 7 | Ivanov 9 | Harvard 9 |
| *Centre* (α, δ) | 06:15:48 14:15:51 | 07:30:39 -15:18:04 | 06:59:47 -04:04:00 | 15:33:44 -53:34:57 |
| *Centre* (G. long., G. lat.) | 196.2134 -1.19776 | 230.9801 1.48416 | 217.4942 -0.01582 | 325.6156 1.89903 |
| R *limited* (arcmin.) | 2.8 | 5.8 | 3.0 | 5.4 |
| Diameter *(pc.)* | 2.36 | 6.96 | 2.18 | 4.55 |
| Core Radius *(arcmin.)* | 0.09 | 0.42 | 0.05 | 0.23 |
| Tidal Radius *(pc.)* | 12.2 | 11.4 | 12.6 | 16.5 |
| Age *(Myr.)* | 5-10 | 400-800 | 80-120 | 200-400 |
| E(J-H) *mag.* | 0.28 | 0.15 | 0.11 | 0.28 |
| E(B-V) *mag.* | 0.90 | 0.48 | 0.35 | 0.90 |
| m-M *mag.* | 11.6 | 12.0 | 10.8 | 11.6 |
| Distance *(pc.)* | 1445±65 | 2065±95 | 1250±60 | 1445±65 |
| $R_{gc}$ *(Kpc)* | 8.60 | 8.65 | 8.23 | 6.10 |
| $X_\odot$ *(pc.)* | 1390 | 1300 | 933 | -1195 |
| $Y_\odot$ *(pc.)* | -405 | -1600 | -760 | -815 |
| $Z_\odot$ *(pc.)* | -30 | 55 | -0.30 | 50 |
| IMF | -2.47 | -2.73 | -3.00 | -2.31 |
| Total Mass ($M_\odot$) | 580 | 470 | 645 | 1450 |
| Members # | 500 | 590 | 665 | 1100 |
| Total luminosity *mag.* | -3.65 | -6.30 | -3.90 | -5.20 |
| Field density (Stars arcmin$^2$) | 4.5 | 7.0 | 4.7 | 11.5 |
| Relax. Time *(Myr.)* | 12.0 | 11.5 | 9.3 | 16.5 |